\documentclass
[twocolumn,english,aps,prl,amssymb,floatfix,groupedaddress]{revtex4}%
\usepackage[T1]{fontenc}
\usepackage{amsmath}
\usepackage{graphicx}
\usepackage{amssymb}
\usepackage{graphicx}
\usepackage{graphicx}
\usepackage{amsfonts}
\usepackage{graphicx}
\usepackage{babel}%
\begin{document}
\title{Self-doping instability of the Wigner-Mott insulator}
\author{S. Pankov and V. Dobrosavljevi\'c}
\affiliation{National High Magnetic Field Laboratory, Florida
State University, Tallahassee, FL 32306}
\date{\today{}}

\begin{abstract}
We present a theory describing the mechanism for the
two-dimensional (2D) metal-insulator transition (MIT) in absence
of disorder. A two-band Hubbard model is introduced, describing
vacancy-interstitial pair excitations within the Wigner crystal.
Kinetic energy gained by delocalizing such excitations is found to
lead to an instability of the insulator to self-doping above a
critical carrier concentration $n=n_{c}$, mapping the problem to a
density-driven Mott MIT. This mechanism provides a natural
explanation of several puzzling experimental features, including
the large effective mass enhancement, the large resistivity drop,
and the large positive magneto-resistance on the metallic side of
the transition. We also present a global phase diagram for the
clean 2D electron gas as a function of $n$ and parallel magnetic
field $B_{\shortparallel}$, which agrees well with experimental
findings in ultra clean samples.

\end{abstract}
\maketitle

Significant experimental advances \cite{kravchenko04} over the last ten years
have provided beautiful and convincing evidence for the existence of a sharp
metal-insulator transition (MIT) in two-dimensional (2D) electron gases
(2DEG). This progress has sparked much renewed interest in better
understanding the basic physical mechanisms that drive the MIT, a fundamental
physical question that has remained poorly understood for many years.

One important issue relates to the stability of an interacting 2D metal with
respect to disorder. While even weak disorder is known to destroy any 2D metal
in absence of interactions \cite{gang4}, very recent work \cite{punnoose02}
has provided strong theoretical evidence that electron-electron interactions
may stabilize a 2D metallic phase. This theory focuses on the most singular
hydrodynamic corrections within the low temperature diffusive regime, and
views disorder as the principal driving force that produces the insulating state.

It should be emphasized, though, that the best evidence for a
sharp MIT is found in the cleanest samples, where the diffusive
regime is restricted to low densities and extremely low
temperatures. Indeed, the experimental data \cite{kravchenko04} \
demonstrating strong effective mass $m^{\ast}$ enhancements have
all been obtained in the ballistic regime, where diffusive
processes are irrelevant. An important question then arises: How
many of the key experimental features can be understood by
deliberately disregarding disorder, and focusing on
\textit{interaction effects alone as the main driving force for
the transition?} This question is the main subject of this paper,
where we propose the quantum melting of a Wigner crystal as the
fundamental mechanism for the MIT in a sufficiently clean 2DEG.

It is well known that at the lowest carrier densities the 2D electrons form a
triangular Wigner lattice. Here, each lattice site is occupied by a single
spin 1/2 electron, since strong on-site Coulomb repulsion prevents double
occupation. The Wigner crystal can therefore be viewed as a (magnetic) Mott
insulator, characterized by an abundance of low-lying spin excitations, but
with an appreciable energy gap to charge excitations. As density increases,
the gap to vacancy-interstitial pair formation decreases 
until the system undergoes a transition to a metallic state. The MIT from such
a Mott insulator to a metal must, therefore, be fundamentally different from
any Anderson-like transition, because the very physical nature of each
insulating state is also completely different. If this idea is indeed correct
- that the Wigner crystal melting is akin to a Mott MIT - then one may expect
the critical behavior of the 2DEG to resemble that of other Mott systems. The
canonical example for the Mott transition in a continuum system is the Fermi
liquid to solid transition in normal $He^{3}$. Here, very recent experiments
\cite{casey03} on 2D monolayers have provided spectacular support for the Mott
transition scenario. The effective mass $m^{\ast}$ was found \cite{casey03}
\ to be strongly enhanced, while the $g$-factor remained essentially
unrenormalized in the critical region - a behavior shockingly similar to that
found in the most recent experiments \cite{kravchenko04} on the 2DEG!

These arguments provide strong motivation to approach the Wigner crystal
melting as a Mott MIT, and develop an appropriate theory for the 2DEG. In the
following, we describe the results of such an approach, demonstrating that the
most striking experimental features of the 2D-MIT can all be simply understood
within this framework.

\emph{Charge transfer model.} 
Our simplified description of a Wigner crystal is based on the idea that there exists a pronounced short range order in charge sector both on the metallic and insulating sides of the MIT. This idea is strongly supported by quantum Monte-Carlo (QMC) work \cite{tanatar89}, which shows that short range charge ordering changes little across the transition. It strongly suggests that, in a ballistic regime, a treatment in terms of an effective lattice model should be applicable on both sides of the transition. The most important elementary excitations across the charge gap of a Wigner crystal correspond to vacancy-interstitial pair formation \cite{lenac95}. It should also be mentioned from the outset that our simplified description does not treat collective modes explicitly, their effect being accounted only through renormalization of effective microscopic parameters of the lattice. 

Sufficiently deep within the insulating phase, the electrons are tightly bound to lattice sites, and such excitations can be well described by an effective charge-transfer (e.g. two-band Hubbard ) model \cite{zaanen85}\ of the form%
\begin{align}
H  &  =\sum_{i\sigma}\varepsilon_{f}f_{i\sigma}^{\dagger}f_{i\sigma}+\varepsilon_{c}c_{i\sigma
}^{\dagger}c_{i\sigma}-\sum_{ij\sigma}t_{ij}c_{i\sigma}^{\dagger}c_{j\sigma
}\nonumber\\
&  +\sum_{i\sigma}V(f_{i\sigma}^{\dagger}c_{i\sigma}+c_{i\sigma}^{\dagger
}f_{i\sigma})+\sum_{i}Uf_{i\uparrow}^{\dagger}f_{i\uparrow}f_{i\downarrow
}^{\dagger}f_{i\downarrow}.\label{hamiltonian}
\end{align}
Here, $f^{\dagger}$, $f$ and $c^{\dagger}$, $c$ are creation and annihilation
operators for site and interstitial electrons respectively, and $U$ is the
on-site repulsion preventing double occupation of lattice sites. In the
tight-binding limit, the band structure parameters $t$ and $V$, and that of
the charge-transfer gap $\Delta_{ct}=\varepsilon_{c}-\varepsilon_{f}$, can be estimated by
computing the appropriate wavefunction overlaps, leading to exponentially
increasing bandwidth with density. The details of such band structure
calculations will not affect any of our qualitative conclusions, and will not
be reported here.

The potentially most serious limitation of our lattice model is its
phenomenological treatment (see below) of elastic deformation (e.g. collective
charge excitations) of the Wigner lattice. These are expected to, at the
least, effectively renormalize \cite{lenac95}\ the band structure parameters,
which should be quantitatively important for the physics of self-doping which
we explore. More importantly, one may question the very justification of using
an effective lattice model, especially on the metallic side of the transition.
There is no particular reason, however, why the suppression of charge ordering
associated with lattice formation must coincides with the closing of the Mott
gap and the MIT. In absence of perfect nesting, the two transitions can occur
separately, and the closing of the Mott gap may be expected to lead to an
itinerant charge-ordered state. At any rate, the existence of short range order on the metallic side justifies, to a significant degree, the use of a lattice model in the inelastic regime. 


\emph{Mott transition via self-doping. }In the Wigner-Mott
insulator, the ground state the electron occupation is naturally
one electron \textit{per unit cell}. However, the lattice
parameters in the 2DEG are self-consistently determined so that,
as the density increases, it may become energetically favorable
for the lattice spacing to slightly expand or contract, while
keeping the charge density \textit{per unit area} fixed (due to
charge neutrality). If this happens, the resulting occupation per
unit cell becomes $1-\delta$, corresponding to an effective
self-doping of our Wigner-Mott insulator. Similar phenomena are
believed to occur near the Mott transition in $He^{3}$
\cite{vollhardt87}, and in the proposed formation of the $He^{4}$
supersolid state \cite{leggett70}. Self-doping may change the
precise nature of the MIT, and thus it needs to be carefully
examined by properly incorporating the electrostatic
considerations that are specific to a charged system of particles.

\emph{Fermi-liquid condensation energy.} Self-doping can be energetically
favorable, since it leads to a kinetic energy gain of delocalized carriers
which condense into a Fermi liquid. The price to pay is the cost of
electrostatic energy to promote a carrier (electron or hole) across the
charge-transfer gap. To assess the stability of the insulator to self-doping,
one must calculate the doping dependence of the condensation energy of the
incipient Fermi liquid state. This requires solving the appropriate Hubbard
model - a general problem where no reliable or accepted theoretical approach
is available at present.

For our purposes a reliable treatment may be possible, and we seek
inspiration from the closely related problem of $He^{3}$
monolayers \cite{casey03}. Here, the observed behavior can be
\textit{quantitatively} understood \cite{casey03} by the simplest
Brinkmann-Rice (BR) theory \cite{vollhardt84} of the Mott
transition. This indicates that one approaches an insulator with
localized magnetic moments (hence a large $m^{\ast}$ as in any
heavy fermion compound), where the inter-site spin correlations
(measured by $g^{\ast}$) in the Mott insulator can be ignored, as
implied by the BR theory. Physically, this may be well justified
for triangular lattices, where both the geometric frustration and
the importance of ring exchange processes \cite{bernu01} conspire
to render the spin correlations negligibly weak in the
experimentally relevant energy (temperature) range.

To implement the BR approach for our problem, we follow the standard methods
\cite{vollhardt84} , where (for simplicity) we have taken $U\rightarrow\infty
$. The free energy (per electron) of the self-doped system then takes the form:%

\begin{multline}
W[\lambda,Z,\mu,\delta]=\\
-\frac{2T}{1-\delta}\sum_{lk}\ln{(1+\exp{(-(E_{lk}-\mu
)/T)})}\\
+\frac{\lambda}{1-\delta}(Z-1)+\mu\label{freeenergy}%
\end{multline}
where $T$ is temperature, ${{E_{lk}}}$ are renormalized band
energies, $\lambda$ is the Lagrange multiplier imposing the occupancy
constraint, $Z$ is the quasiparticle weight , and $\mu$ is the chemical
potential. The free energy $W[\lambda,Z,\mu,\delta]$ is stationary in the
ground state: $\partial W/\partial a=0$, where $a=\lambda,Z,\mu,\delta.$

The two bands of the model are coupled via the hybridization $\sqrt{Z}V$. We
further assume that the interstitial band density of states (DOS) is
approximated by a regular function $\nu(\epsilon)$.
The renormalized band energies then explicitly read :%

\begin{multline}
E_{1,2}(\epsilon)= \frac{1}{2}(\varepsilon_{f}+\varepsilon_{c}+\epsilon+\lambda)\\
\mp\frac{1}{2}\sqrt{(\varepsilon_{c}-\varepsilon_{f}+\epsilon-\lambda)^{2}+4 Z V^{2}}
\label{bands}%
\end{multline}

The hybridization $V$ and the density of states $\nu(\epsilon)$ have explicit
dependence on the lattice spacing and hence on the doping $\delta$. The
density depends on $\delta$ as $\nu(\epsilon,\delta)=\nu(\epsilon
/\gamma(\delta))/\gamma(\delta)$, where $\gamma(\delta=0)=1$. The details of
the dependence of $V$ and $\gamma$ on $\delta$ are not of qualitative
importance near the transition, as long as $V(\delta)$ and $\gamma(\delta)$
are smooth functions of $\delta$. Note that the van Hove singularities of a triangular lattice are sufficiently far from the Fermi energy at half filling.

The choice of $\varepsilon_{f}$ and $\varepsilon_{c}$, on the contrary, proves very important. We
use effective electrostatic energy parameters to model complex energetics of
the problem, arising from strong renormalization of bare parameters by elastic
modes. For simplicity we assume a linear dependence of local potentials on the
charge densities of the site and interstitial sublattices:%

\begin{equation}
\varepsilon_{i}=\frac{v_{ij}}{\sqrt{1-\delta}}n_{j} \label{ei}%
\end{equation}
where $\{i,j\}=\{f,c\}$, $n_{f}=1-Z$ and $n_{c}=Z-\delta$, so the effective
potentials depend on both $\delta$ and $Z$. The prefactor $1/\sqrt{1-\delta}$
represents overall rescaling of the Coulomb interaction with the change of the
lattice spacing, due to charge neutrality. The coefficient $v_{ij}$ is an
effective potential created on an (empty) site of the band $i$ by the fully charged band $j$. We expect the polarization to play a crucial role in the potential
renormalization. When a hole is created the nearby electrons are attracted to
the vacancy, partially screening it. When an electron is placed in the
interstitial orbital, the nearby electrons are repelled \cite{lenac95}, again
partially screening the charge fluctuation. The effect of the screening is
always directed toward decreasing the energy of the corresponding
particle-hole excitation. Due to the elastic softness of the Wigner lattice
(e.g. shear phonons with the energy two orders of magnitude smaller, than the
bare Coulomb energy) we expect a strong renormalization of the excitation
energy, leading to an appreciable reduction in of the charge-transfer gap
$\varepsilon_{c}-\varepsilon_{f}$ . Therefore we assume that $(v_{ff}-v_{cf})/v_{ff}=\alpha\ll1$.
The value of $v_{cc}$ enters only the second order corrections in $\delta$, so
we ignore it and simply set to zero. The stability requirement for the
classical Wigner crystal restricts the value of $v_{fc}$. The electrostatic
energy of the WC is $E=(\varepsilon_{f}(1-Z)+\varepsilon_{c}(Z-\delta))/(1-\delta)$. In the
classical limit $Z=\delta$ for $\delta>0$ and $Z=0$ for $\delta<0$. By setting
$v_{cf}=(1-\alpha)v_{ff}$ we find that the WC is stable only for
$v_{fc}>v_{ff}(1/2+\alpha)$.
\begin{figure}[ptb]
\vspace{.0cm} \includegraphics[width=3.0in]{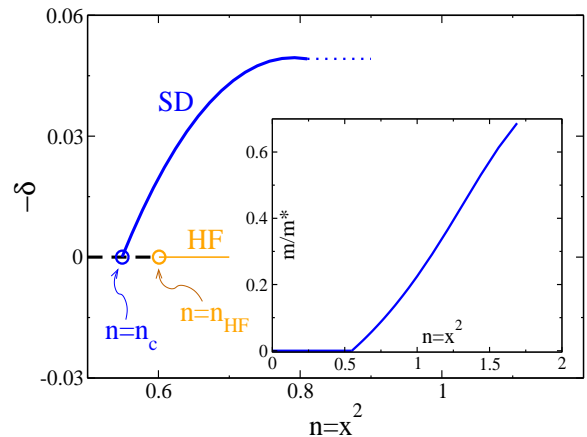} \vspace
{.5cm}\caption{(Color online) The phase diagram in the zero magnetic field. The MIT takes route via self-doping (thick solid line), which always precipitates the transition at half-filling (thin solid line). The inset shows the quasiparticle weight, vanishing at $n_c$.}%
\end{figure}

\emph{Solution of the model.} The problem can be simplified in the critical
regime (when $Z\rightarrow0$ and $\delta\rightarrow0$) and analyzed
analytically. This allows us to make certain general statement about the
nature of the MIT. Away from the transition we resort to a numerical solution.

\emph{a. Linear analysis and stabilization of the metallic phase via
self-doping.} In the following we show that the self-doping (SD) transition
always precipitates the transition taking place at half-filling (HF). We
obtain a criterion for determining whether the SD is electron or hole like.

We expand our equations around the insulating solution ($Z=0$,
$\delta=0$) to linear order in variations of the parameters
$\lambda$, $Z$, $\mu$, $\delta$, assuming $T=0$. At the transition
point $\mu=\varepsilon_{f}+\lambda$ and the free energy is purely classical:
$W_{c}=\varepsilon_{f}$. From the saddle point equation $\partial
W/\partial\delta=0$ we find that in the SD transition $W-\mu
=-\partial \varepsilon_{f}/\partial\delta$, hence $\lambda_{c}=\partial \varepsilon_{f}%
/\partial\delta$.

The results of the expansion in small $Z$ can be conveniently summarized in
terms of the axillary functions $\phi_{1}(V,\lambda)$ and $\phi_{2}%
(V,\lambda)$ defined as:
\begin{align}
&  \phi_{1}(V,\lambda)=\lambda+\frac{\partial \varepsilon_{f}}{\partial Z} -2V^{2}%
\int_{-\infty}^{E_F} d\epsilon\nu(\epsilon) \frac{1}{\varepsilon_{c}-\varepsilon_{f}+\epsilon
-\lambda}\label{phi1}\\
&  \phi_{2}(V,\lambda)=1-2V^{2}\int_{-\infty}^{E_F} d\epsilon\nu(\epsilon)
\frac{1}{(\varepsilon_{c}-\varepsilon_{f}+\epsilon-\lambda)^{2}} \label{phi2}%
\end{align}
where $E_F$ is the Fermi energy.
One can check that the equation $\partial W/\partial Z=0$ for $Z\to0$ (at SD
or HF transition) is simply $\phi_{1}(V,\lambda)=0$. Differentiating this
equation (where $V=V(\lambda)$ satisfies the equation) with respect to
$\lambda$ we have:%

\begin{equation}
0=\frac{d\phi_{1}}{d\lambda}=\phi_{2}-2\frac{\lambda+\frac{\partial \varepsilon_{f}%
}{\partial Z}}{V}\frac{dV}{d\lambda} \label{dphi1dlambda}%
\end{equation}

It follows from the saddle point equations $\partial W/\partial\lambda=0$ and
$\partial W/\partial\mu=0$
that at the HF transition $\phi_{2}(V,\lambda)=0$. Therefore, according to the
Eq.(\ref{dphi1dlambda}), $dV/d\lambda=0$ at the HF transition. Direct
inspection indicates that $W$ has a maximum there, thus the SD transition
always occurs before the HF transition.

By considering $\partial W/\partial\lambda=\partial W/\partial\mu=0$ near the
SD transition one finds that $Z=\delta/\phi_{2}(V,\lambda_{c})$. Therefore, if
$\phi_{2}$ is positive in the SD transition, then the doping is hole-like, and
if $\phi_{2}$ is negative the doping is electron-like. If $\phi_{2}=0$ the SD
transition coincides with the MIT transition restricted to half filling.

\emph{b. Numerical solution.} We choose the parameters of the model that can
best mimic the experimental results. For that we set $v_{ff}=E_{C}$,
$v_{cf}=(1-\alpha) E_{C}$, $v_{fc}=0.1 E_{C}$ where $\alpha=0.1$ and $E_{C}$
is the bare Coulomb energy. We use a parameter $x=D_{c}/|E_{C}|$ to mimic the
$r_{s}$ number, where $D_{c}$ is the width of the interstitial band. The
electron density goes as $n\sim x^{2}$ in a 2D electron gas. We set $V=D_{c}$.
The density of states in the interstitial band is constant, mimicking a two
dimensional dispersion. For these choice of the parameters we find
(Eqs.(\ref{phi1},\ref{phi2})) that the SD transition occurs at $x_{SD}=0.7408$
($n_{SD}=.549$) and the HF transition occurs at $x_{HF}=0.7751$ ($n_{SD}=.601$).

In contrast to standard Mott transition, the half-filled insulator (heavy
dashed line in Fig. 1) thus becomes unstable to electron-like self-doping
(heavy full line in Fig. 1),\textit{ before the half filled transition} takes
place (thin full line in Fig. 1). The quasi-particle weight $Z\sim1/m^{\ast
}\sim(n-n_{c})$ vanishes linearly (see inset of Fig. 1) as the transition is
approached from the metallic side, in agreement with experiments
\cite{kravchenko04}.

\emph{Transport, effect of magnetic field, and phase diagram. }
Different properties has been studied in detail for various Mott systems using recently developed DMFT method \cite{dmft} - a reliable tool even at low dimensions (unless the critical properties are specifically tied to the system's dimensionality).
This approach can be regarded as a finite temperature generalization of the BR
theory we utilized. Armed with this knowledge, one can directly list what is
expected within the framework we consider: (1) Below the transition, transport
takes place by activation $\rho(T)\sim\exp{(-\Delta}_{o}{(n)/T)}$, with
$\Delta_{o}(n)\propto n_{c}-n$, just as seen in the experiments
\cite{kravchenko04}; (2) On the metallic side, heavy quasiparticles exist only
below a coherence temperature $T^{\ast}(n)\sim1/m^{\ast}\sim(n-n_{c})$,
leading \cite{aguiar05} to a large resistivity drop \cite{kravchenko04} at
$T<T^{\ast}(n)$; (3) A parallel magnetic field $B_{\shortparallel}^{\ast
}(n)\sim1/m^{\ast}$ $\sim(n-n_{c})$ is sufficient to produce full
spin-polarization of the electrons, destroying the coherent quasiparticles and
causing large and positive magneto-resistance \cite{kravchenko04}; (4) Close
to the transition, at $B_{\shortparallel}>B_{\shortparallel}^{\ast}(n)$ the
resistivity saturates to a field-independent value $\rho(T)\longrightarrow
\rho_{\infty}(T)$, which assumes an activated form $\rho_{\infty}(T)\sim
\exp{(-\Delta}_{{\infty}}{(n)/T)}$, where the gap ${\Delta}_{{\infty}}{(n)}$
remains finite in the high field limit. This behavior is specific to a
charge-transfer (CT) model we consider, since the charge transfer gap
$\Delta_{ct}$ remains finite as $B_{\shortparallel}\longrightarrow\infty$, in
contrast to the standard Mott gap $\Delta_{Mott}=U+g\mu_{B}B_{\shortparallel}%
$. (5) In the CT model the MIT reduces to a band-crossing transition in the
$B_{\shortparallel}\longrightarrow\infty$ limit, where ${\Delta}_{{\infty}%
}{(n)}\sim(n_{c}^{\infty}-n)$ vanishes at $n_{c}^{\infty}>n_{c}$,
and the system remains metallic at higher densities. (6) The
resulting phase diagram (Fig. 2) agrees well with the experimental
one \cite{jaroszynski04} obtained for ultra-clean samples.
\begin{figure}[ptb]
\vspace{.0cm} \includegraphics[width=3.0in]{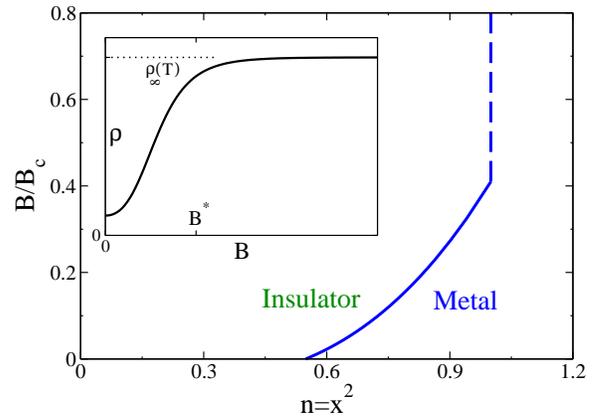} \vspace
{.5cm}\caption{The metal-insulator phase diagram in the presence of a parallel
magnetic field (in units of $B_c=n_c/\mu_Bgm$).
The dashed line represents a metal to band insulator
transition. The electron spin becomes fully polarized at the magnetic field
$B^{\ast}$. The inset shows the change in resistivity from Fermi liquid to
insulating behavior as the magnetic field exceeds $B^{\ast}$.}%
\end{figure}

\textit{Conclusions. }We presented a theory for the
interaction-driven MIT describing the clean 2DEG. Our approach
focuses on vacancy-interstitial excitations within a Wigner-Mott
insulator, naturally leading to a two-band (charge transfer)
Hubbard model. As density increases, such excitations lead to an
instability of the insulating state, and produce a self-doping
driven Mott transition to a heavy electron metal. The general
predictions of this model seem to explain most puzzling features
seen in the experiment, strongly suggesting that Coulomb
interactions and not disorder provide the fundamental driving
force for the 2D-MIT.

The most challenging task for future work is to extend the present
approaches to explicitly include the dynamics of collective charge
fluctuations which are phenomenologically treated in the
considered lattice model. This goal should be facilitated by
recent advances\ \cite{pankov05} in theories for Coulomb gap
phenomena, and would provide a more rigorous justification of the
lattice model we introduced. Even more importantly, such a theory
will be indispensable to understand experiments \cite{tsui06} at
temperatures and densities where the Wigner lattice has already
melted, but where strong short-range charge correlations persist.
Such a regime is of appreciable importance and extent whenever the
reduced Coulomb interaction strength $r_{s}\gg1$, as found in many
experiments on the 2DEG.

We acknowledge fruitful discussions with E. Abrahams, G. Kotliar,
E. Manousakis, D. Morr, A. Punnoose, and J. Schmalian. This work
was supported through the NSF Grant No. DMR-0542026 (V. D.) and
the National High Magnetic Field Laboratory (S. P. and V. D.). We
also thank the Aspen Center for Physics, where part of this work
was carried out.

\end{document}